\documentclass[letterpaper,12pt,thesis]{SeniorProject}

\usepackage[normalem]{ulem}	
\usepackage{multirow,	
	graphicx,
	units,		
	cite,			
	amsmath,
	enumerate,	
	amssymb,
	array,		
	calc,			
	}
\usepackage[colorlinks=true,linkcolor=blue]{hyperref}

\begin{document}

\title{Teaching practices of graduate teaching assistants}
\author{Eric Matthew Hickok}
\majorfield{DEPARTMENT OF PHYSICS AND ASTRONOMY} \depositdate{2015}

\begin{preface}



\begin{singlespace}
	\begingroup
		\hypersetup{linkcolor=black}		
		\tableofcontents
		\listoftables
		\listoffigures
	\endgroup
\end{singlespace}

\prefacesection{Abstract}

	Physics education research has consistently shown that students have higher learning outcomes when enrolled in interactive-engagement courses. Consequently, many schools are actively reforming their introductory curricula. For courses where the interactive sections (labs, tutorials, and/or workshops) are mostly taught by graduate student teaching assistants (TAs), good TAs are instrumental to the success of the reform. Many studies have investigated specific interactions between TAs and students, but more can be learned through a holistic examination of TA-student interactions. Over the course of one semester, I observed TAs in their various teaching roles using the Real-time Instructor Observation Tool (RIOT). These observations serve to show what TAs may ``default to'' with little to no intervention. I present a snapshot of a department in the early stages of reform and discuss the implications of my findings as they relate to the professional development of our TAs.

\prefacesection{Introduction}

A common theme in physics education research is the overwhelming evidence that students tend to learn more when they are actively engaged in the learning process \cite{hake,
		mazur}. Students enrolled in interactive-engagement (IE) courses show greater normalized gains in conceptual understanding (as measured by tests like the {\em Force Concept Inventory} \cite{fci}) than their peers enrolled in traditional courses \cite{hake}. Arguments have been made that IE students have better retention of material \cite{Francis1998Do} while achieving the same, if not better, problem solving skills \cite{Hoellwarth2005Direct}. Indeed, the results have been so convincing that high schools and universities across the country have already adopted, or are moving to adopt, their own version of an IE course \cite{Finkelstein:Replicating,
		Potter:Sixteen,
		Goertzen2011Moving, 
		Kohl2012Chronicling, 
		Hestenes2011Graduate, 
		mazur, 
		McDermott2000Preparing}.
	
The Department of Physics and Astronomy at San Jos\'{e} State University (SJSU) is engaged in the preliminary stages of adopting an IE curriculum for its introductory physics courses. Like most universities, many of SJSU's laboratory sections of these courses are currently taught by graduate student teaching assistants (TAs). Being a terminal master's program, the pool of available TAs is generally in flux, providing approximately three years of teaching experience for any given TA. Although much has been studied examining TAs in an IE environment \cite{koenig, 
		Goertzen2008Indicators,
		Goertzen2009Accounting, 
		Goertzen2010Tutorial
		} there has been little study of graduate students in a laboratory teaching environment such as ours. Additionally, while much has been written about how best to train and prepare TAs for their teaching duties \cite{Robinson2000New,
		Luft2004Growing,
		Price2008Preparing,
		Goertzen2009Accounting,
		Spike2012Preparing
		}, remarkably little research has been to see exactly what they are in fact doing during their teaching assignments in more traditional laboratory settings.
		
Much work has been done studying effective methods of preparation for TAs (and undergraduate learning assistants) in a Studio Physics environment \cite{Price2008Preparing,
		Spike.2011.Toward-PERC.PeerReviewed.363,
		Spike2012Preparing
		}. However, our labs are not of the ``studio'' nature. In the case of our calculus-based freshman mechanics course, the lab section is largely traditional. Students work from a lab manual with a prescribed procedure, the result of which ``validates'' concepts or values described or derived in lecture. Students also attend a weekly problem-solving session. These ``workshops'' are meant to be time for students to work collaboratively in small groups to solve end-of-the-chapter-style physics problems.
		
Study of lab and workshop sections of this introductory course shows the state of a department in the very early stages of reform. The aim of this project is to develop this baseline of data showing what our TAs may ``default to'' with little to no intervention. This information will be analyzed by class format and then be compared to best practices from current literature.
	
\end{preface}

\chapter{Background}
	\section{Motivation}
Some teachers advocate for the practice of asking students how they think they learn best and trying to adapt their teaching methods to suit student preferences. Unfortunately, students tend to prefer a passive learning environment and even think they learn better in it \cite{koenig}. 
Despite this preference, Koenig, {\em et al.} found that students performed significantly better on post-tests in an IE setting lead by a TA than in a lecture delivered by a professor. This result is so prevalent that the majority of the literature involving the study of TAs occurs in an IE setting \cite{Goertzen2008Indicators,
		Gray.2009.Analysis-PERC.PeerReviewed.149,
		Spike.2010.Examining-PERC.PeerReviewed.309,
		Spike.2011.Toward-PERC.PeerReviewed.363,
		Goertzen2011Moving
		}. Despite varying beliefs about teaching, TAs tend toward similar teaching behaviors in the classroom \cite{Goertzen2010Tutorial}. 

Goertzen, {\em et al.} suggest that TA ``buy-in'' to reformed instruction is a necessary component of effective curriculum implementation and that detailed understanding of TA beliefs and motivations can only improve TA professional development \cite{Goertzen2009Accounting}. It is important for any department employing TAs, regardless of what stage of reform they are in, to tailor their professional development to meet the needs of their TAs. Providing TAs the opportunity to express their beliefs and opinions about teaching allows for responsive professional development \cite{Goertzen2010Respecting}. Much as we hope to build on productive seeds of physics knowledge in our students, we should be observing our TAs' teaching so that we can see and encourage beliefs and practices that are more aligned with reformed physics instruction.

Studies have found considerable variation in how TAs interact with their students~\cite{Goertzen2009Accounting, wilcox}, even when the TAs are provided with extensive notes about the curriculum~\cite{West2013Variation}. Even different instructors' implementations of the same reform strategies have shown striking differences~\cite{Turpen2009All,
		Turpen2010Construction
		}. Not only is it important then to observe the TAs to improve professional development offerings, but also to examine how a curriculum is being interpreted by the TAs and deployed to the students. Observations of TAs provides an opportunity for the department to examine curriculum and make any needed improvements, from use of equipment to lessons as a whole.

	\section{Guiding questions}	
		\label{sec:guidingqs}
For courses where the interactive sections (labs, tutorials, and/or workshops) are mostly taught by graduate student teaching assistants, good TAs are instrumental to the success of the reform \cite{koenig}. Consequently, TA professional development is crucial to successful reform.

Before developing a robust professional development program that is targeted to the needs of our TAs, we first need to know where they are starting out. Therefore, it is imperative to observe their interactions in their teaching assignments. These observations will serve to show what practices TAs may ``default to'' with little to no intervention, providing a snapshot of the department in the early stages of reform.

This study was envisioned to answer the following questions.
	\begin{itemize}
		\item What teaching practices are currently being used by our TAs?
		\item How are the current teaching practices split by format?
		\item How do these practices compare to what is currently being done by similar departments and institutions?
		\item How do these practices compare to what physics education research tells us are the most effective practices?
	\end{itemize}
I will compare the teaching practices in use in our different lab and workshop courses. Answers to these questions will provide a baseline of information to further the ongoing conversation in the department about improving instruction in the introductory physics series.

	\section{Observational tools}
		\label{sec:tools}
		Education research has yielded a number of tools for measuring the degree to which a course or instructor utilizes interactive and student-centered techniques. One such tool for this is the Reformed Teaching Observation Protocol (RTOP) \cite{rtop}. 
	
With sufficient training, an RTOP observer rates the instruction of a given lesson on the topics of lesson design, content, and classroom culture using a Likert Scale. In our case, the TAs teach from a prescribed lab manual without the freedom to design the labs according to the specific needs of their students. Only the prompts concerning classroom culture would provide insight into the TA's role in the classroom. These prompts account for only ten of the twenty-five items, with only five directly concerning TA-student interactions. How TAs choose to interact with their students is the most overt choice of teaching practice available to them. And while the RTOP encourages documenting events to justify the ratings, as a tool, it doesn't provide the level of resolution for our needs, with only five of the twenty-five items offering insight into TA-student interactions.

For our needs, a more useful observational protocol would provide detail into how TAs and students are spending their time in the classroom without judging the effectiveness of any specific pedagogic strategy. A team at the University of Maine and the University of British Columbia developed the Classroom Observation Protocol for Undergraduate STEM (COPUS) to do just this \cite{copus}. The COPUS was developed as a modified version of the Teaching Dimensions Observation Protocol (TDOP) \cite{tdop}. The team found that the TDOP required too much training and offered too many codes to fit their specific needs. The COPUS is simplified to code what both the students and the instructor are doing at two-minute intervals into 25 possible activities and interactions, twelve for students and thirteen for instructors. Completed analysis of an observation yields two pie charts showing the prevalence of each code for both students and instructors. The individual codes of the COPUS reveal that the protocol is designed to code nonlaboratory-style courses. The majority of the codes would be of little use in a laboratory or workshop setting.

Researchers at UC Davis originally developed a tool for monitoring and classifying instructor-student interactions \cite{West2013Variation}. Later made publicly available by a team at San Jos\'{e} State University, this Real-time Instructor Observing Tool (RIOT) quantifies and times these interactions, providing a more illustrative view of what actually takes place in a given TA's classroom. The RIOT enables categorization and recording of instructor actions continuously in real time. Interactions are categorized into four major groups: talking at students, shared instructor-student dialogue, observing students, and not interacting. Each of these major categories contains useful subdivisions of several descriptive categories as shown on \hyperref[tab:riot]{Table \ref*{tab:riot}}, adapted from West, {\em et al}.\ 2013 \cite{West2013Variation}. The instructor can perform each subcategory of interaction with an individual student, a small group of students, or with the whole class. Interaction-type, start time of the interaction, and duration of the interaction are all logged in real time, providing a full picture of TA-student interactions for the duration of the class period.

	\newcommand{\leftcol}{0.2\textwidth-2\tabcolsep}
\newcommand{\midcol}{0.24\textwidth-2\tabcolsep}
\newcommand{\rightcol}{\textwidth-\leftcol-\midcol+2\tabcolsep}

\begin{table*}
	\centering
	\caption[A list of Instructor-Student interactions captured by the RIOT]{A list of all possible Instructor-Student interactions captured by the RIOT, adapted from West, {\em et al}.\ 2013 \cite{West2013Variation}.}
	\label{tab:riot}
	\begin{tabular}{
		p{\leftcol}
		>{\raggedright\arraybackslash}p{\midcol}
		p{\rightcol}
		}
		\hline\hline
		\parbox[t]{\leftcol}{Type of \\Interaction}	&	Category of Interaction	&	\parbox[s]{\rightcol}{Description (Instructor is\ldots)}	\\\hline
		\multirow{2}{*}{\parbox[t]{\leftcol}{Talking \\At Students}}	&	Clarifying Instructions	&	Clarifying the instructions, reading from the activity sheet, covering logistical issues, transitioning, \ldots.	\\\cline{2-3}
		&	Explaining Content	&	Explaining physics concepts, answers, or processes to student(s).
			\\\hline	
			
		\multirow{4}{*}{\parbox[t]{\leftcol}{Dialoguing with Students}}	&	Listening to Question	&	Listening to a student's question.\\\cline{2-3}
			&	Engaging in Closed Dialogue	&	Asking a series of short questions meant to lead the student to a correct answer. Student contribution is one to several words at a time.\\\cline{2-3}
			&	Engaging in Open Dialogue	&	Students are contributing complete sentences, though not actively ``making sense.''\\\cline{2-3}
			&	Ideas being shared	&	Participating in student-led conversation. Student contribution is complete sentences with concepts being challenged and worked on.\\\hline
			
		\multirow{4}{*}{\parbox[t]{\leftcol}{Observing Students}}	&	Passive Observing	&	Scanning room and assessing student progress from afar or browsing whiteboard work of groups for less than ten~seconds at a time.\\\cline{2-3}
			&	Active Observing	&	Actively listening to small groups or individuals.\\\cline{2-3}
			&	Students Presenting	&	Listening to students presenting their work to the class.\\\cline{2-3}
			&	Students Talking Serially	&	Listening to students talking serially, asking each other questions and building on each others' ideas.
			\\\hline
			
		\multirow{4}{*}{\parbox[t]{\leftcol}{Not \\Interacting}}	&	Administrative and/or Grading	&	Grading student homework, or discussing quizzes or other course policies.\\\cline{2-3}
			&	Class Preparation or Reading TA Notes	&	Reading notes, or writing something on the board.
			\\\cline{2-3}
			&	Chatting	&	Chatting socially with students. This is not an interaction concerning physics.\\\cline{2-3}
			&	Working on Apparatus and/or Material	&	Helping students with experimental apparatus or computers. Any possible discussion is devoid of any physics content.\\\cline{2-3}
			&	Out of room	&	Left the room.\\
			\hline\hline
	\end{tabular}
\end{table*}


	\section{Sample and Environment}
		\label{sec:sample}
		For this study, observations took place in two distinct TA-led course types: introductory physics labs and introductory physics workshops. The lab sections are fairly traditional, with students following a prescribed procedure to achieve a result that is then compared with the theory developed in the lecture course. Workshops were instituted to provide students an opportunity to work in small groups to solve end-of-chapter style problems and present their solutions to their classmates.

Learning outcomes for the laboratory section include the ability to ``solve relatively complex mechanics problems in a systematic manner by applying the laws of physics and calculus'' and ``qualitatively describe the motion of objects using physics terminology and concepts.'' Each lab session meets once a week for two hours and fifty minutes.  In lab, students work in small groups to complete an experiment and report their findings. Some of the courses require students to take periodic quizzes during lab time to make sure they are keeping up with the material. Labs are taught by either faculty or graduate TAs. The lab instructor is responsible for providing brief background information the day's activity, demonstrating the materials to be used as needed, troubleshooting student difficulties and finicky equipment, and grading lab reports and any quizzes. Laboratory sections are limited to 20 students per section.
	
The introductory, calculus-based mechanics course at SJSU (Physics~50) has an optional-but-strongly-encouraged ``workshop'' component (Physics~50W). The workshops are modeled after workshops that have been used successfully in the Calculus~1 courses at SJSU as well as in physics courses at other universities. Students taking those workshops experienced a significant rise in course grades, and a significant decrease in the probability of failing, compared to those who did not take the workshops.
	
Workshops meet one a week for one hour and fifty minutes. Students work in small groups to complete a packet of physics problems. The intention is that students work together collaboratively to solve these problems on whiteboards, taking turns solving the problems and leading small-group discussions. The role of the TA in workshop is that of a facilitator: take attendance and help students solve the problem sets through hints and guiding questions. The TAs are provided with solutions to the problem sets, but problems are not graded. The TA merely notes that groups are making satisfactory progress. Workshops are limited to 24 students per section.
	
Roughly one-third of the Physics~50 lab meetings consisted of ``problem solving sessions'' wherein students were given a packet of physics problems to solve rather completing a laboratory exercise. These problem solving sessions are meant to be run in a similar manner to the workshop sessions. Students work with the usual lab group, working through a packet of physics problems. Typically, one or two problems of the set are turned in to be graded while the rest are offered simple as additional practice.
	
In order to teach a lab or workshop section, prospective TAs must complete an application with the department. They must also be a current Masters student in the Department of Physics and Astronomy with classified standing, must have an overall GPA of 3.0 or higher, and must be enrolled in three units or more at SJSU. Prior to the first day of classes, TAs attend a one-day training touching on classroom management and grading policies as well as a brief overview of physics education research findings regarding pedagogy. During the training, TAs are instructed to ask the students leading questions about their reasoning rather than provide them with solutions or merely correct answers. TAs are assigned sections to teach according to departmental need and their own course schedule, at the discretion of the department chair.
	
The TAs that were observed were in their first or second year of graduate school. For the most part, their teaching experience was limited to tutoring and previous lab or workshop teaching at SJSU.
	
As is common in many universities, students register for separate lecture and lab sections. A given lab section can therefore have students from multiple lecture sections taught by multiple lecture instructors. In the interest of consistency, TAs teach from a prescribed Lab Manual so that every lab section of a given course conducts the same experiment each week. Thus, some students entering their lab course on a given day may not have seen the material in their lecture section while other students may have. It is left to the TA to describe the requisite material for the day's activity as well as teach the students how to safely use the equipment to make adequate measurements. The TAs are also responsible for teaching sufficient data analysis methods for the students to make sense of the data they collect. This is often taught directly, in the form of a short lecture. During the course of the semester, TAs are observed at least once by the professor in charge of a given course's lab sections and provided with written feedback about their teaching.
		
	\section{Targeted questions}
		Working within the limitations of the RIOT, teaching assignments, and observation scheduling, the guiding questions laid out in \hyperref[sec:guidingqs]{Section~\ref*{sec:guidingqs}} require some refinement.

The choice of the RIOT as my observational tool limits my observations to only things done by or with the TAs. My observations will completely bypass any work done by the students that does not directly involve the TA. It comes to a question of the intent of the course or TA: If the course is designed such that students work quietly in groups with little to no intervention from the TA (or if the TA structures the learning environment this way), the only way to capture this using the RIOT is through notes. The RIOT would log this time as {\em Not Interacting} for the TA even though the students are actively working and possibly learning. With that in mind, I can only effectively log the TA-student interactions listed in \hyperref[tab:riot]{Table~\ref*{tab:riot}}. This leads to the first question I intend this study to address.

\begin{itemize}
	\item[1.] What TA-student interactions (as identified by the RIOT) do our TAs employ in their teaching assignments?
\end{itemize}

For the semester in question, Spring 2015, the majority of our TAs were assigned to teach at least the laboratory section of Physics~50 or the workshop section, Physics~50W. Some were also assigned to the laboratory sections of the algebra-based freshman mechanics course or the calculus-based freshmen E\&M course, Physics~2A and Physics~51 respectively.

My own schedule, coupled with the teaching assignments, allowed me to make very few observations in Physics~2A and Physics~51. These two courses are included in my overall results for completeness in answering the above question. Lacking a significant number of observations, I instead focused my analysis on the physics courses ``for majors.'' Having observed the majority of the TAs assigned to both Physics~50 and 50W, it is apposite to examine these courses in more detail, leading us to ask the following.

\begin{itemize}
	\item[2.] How do these interactions differ in Physics~50 and Physics~50W?
\end{itemize}

A small number of my observations in Physics~50 were of ``problem solving days.'' Considering that on these days students were not working an actual experiment but rather working problem sets, one might assume these observations to be indistinguishable from an observation of Physics~50W. A subquestion to Question 2 is worth consideration.

\begin{itemize}
	\item[2a.] How do Physics~50 ``problem solving days'' compare to regular lab days and Physics~50W observations?
\end{itemize}

Research using the RIOT is not yet widespread. Still we can compare our results to similar studies at other institutions to gain insight into the level of interactivity our labs and workshops provide.

\begin{itemize}
	\item[3.] How do our TAs' implementations of Physics~50 and Physics~50W compare with available RIOT data from courses at UC~Davis?
\end{itemize}

As instructors, we should always strive to make our courses the most effective they can be. With that in mind, we ask the following question.
\begin{itemize}
	\item[4.] How do our TAs' implementations of Physics~50 and Physics~50W compare with PER-based best practices?
\end{itemize}

\chapter{Methods}
	
	\section{Observations}
		This study focused on the interactive nature of the introductory physics laboratories and workshops taught by TAs at SJSU. Observations took place throughout the spring semester of 2015. Of the eight physics TAs employed by the department, I observed a total of five. There were a total of sixteen laboratory or workshop sections taught by TAs of which I observed nine. Each section was observed at least twice.

Eight sections of the Physics~50 lab were taught by a total of five TAs. I observed three of these TAs teaching four of these sections. Five sections of the Physics~50 Workshop were taught by four TAs. I observed three of these TAs teaching three of these sections. Only one TA was observed in both Physics~50 and 50W. A breakdown of these numbers is shown in \hyperref[tab:obs_counts]{Table~\ref*{tab:obs_counts}}.

\begin{table*}[ht]
	\centering
	\caption{Observation counts by TA and course}
	\label{tab:obs_counts}
	\begin{tabular}{rcccccc}	
		\hline\hline
		• & \multicolumn{2}{c}{Physics~50} & \multicolumn{2}{c}{Physics~50W} & \multicolumn{2}{c}{Total$^\dagger$} \\ 
		• & Observed & Out of & Observed & Out of & Observed & Out of \\ 
		\hline
		TAs & 3 & 5 & 3 & 4 & 5 & 8 \\ 
		Sections & 4 & 8 & 3 & 5 & 9 & 16 \\ 
		\hline\hline
	\end{tabular}\\
	\begin{tabular}{rp{4.75in}}
	$^\dagger$	&	{\footnotesize It should be noted that I observed a single TA teaching the Physics~2A lab and another teaching the Physics~51 lab. To maintain anonymity, these observations are only included here and in the overall results (\hyperref[sec:overall_results]{Section~\ref*{sec:overall_results}}) and not discussed on their own. Also, there was some overlap by individual TAs teaching multiple courses so the ``Total'' numbers are not the sums of the course numbers.}
	\end{tabular}
\end{table*}
	
Observations took place for the first hour of instruction. For those sessions that a quiz was administered at the beginning of the class, I extended my observation to include the hour of teaching time after the quiz had taken place. Aside from introducing myself and describing why I was present when asked, I did not interact with the instructor or the student during the course of the observation and no feedback was provided to the instructors about their teaching practice until the end of the semester when all my observations had been completed.
	
To maintain anonymity, each TA was assigned a pseudonym and all data were logged under these pseudonyms. I coded the observations using the RIOT.
		
	\section{Validity of measurements}
		Before beginning my official observations, I observed three classes with another researcher familiar with the RIOT. After each of the three sessions, we discussed the differences in our observations at length resulting in a near match for our third observation. After the third session, our observations showed an agreement of 88.5\%. Accounting for chance agreements, we saw a Cohen's kappa value of 82.5\% \cite{cohen_1960_coefficient}. Even by conservative measures, these values are sufficient to establish the validity of my measurements.
		
		\subsection{Cohen's kappa}
		To determine these values, I wrote a python script to compare observations of a single class session taken by two different observers. For each observed interaction, the RIOT notes the start time and a duration, each with one second resolution. Using these, my script builds a time series of each observation show the clock time and interaction for each second of the observation. It then compares the two observations to find temporal overlap in the two. This way, the observers don't need to exactly synchronize the first interactions noted by their RIOT instance.

With a common start time established, the script goes through every second of the observations and constructs a matrix of what interaction each observer marked for every second, an example of which is shown in \hyperref[tab:kappa_matrix]{Table~\ref*{tab:kappa_matrix}}. 
\begin{table*}[ht]
	\centering
	\caption[A sample matrix used to calculate Cohen's kappa]{A sample matrix used to calculate Cohen's kappa. The ``22'' value in the top right indicates that Observer A selected {\em Not Interacting} while Observer B had selected {\em Clarifying Instruction} a total of twenty-two times during the observation in question.}
	\label{tab:kappa_matrix}
	\begin{tabular}{cr|cccccc}
	\hline\hline
	& & \multicolumn{6}{c}{Observer A} \\
	& & {\rotatebox[origin=l]{90}{Clarifying}} & {\rotatebox[origin=l]{90}{Explaining}} & {\rotatebox[origin=l]{90}{Listening}} & {\rotatebox[origin=l]{90}{Closed Dialogue}} & $\cdots$ & {\rotatebox[origin=l]{90}{Not Interacting}} \\
	\hline
	\multirow{6}{*}{\rotatebox[origin=c]{90}{Observer B}}& Clarifying & 253 & 3 & 12 & 0 & $\cdots$ & 22 \\
	& Explaining & 5 & 658 & 0 & 12 & $\cdots$ & 6 \\
	& Listening & 21 & 48 & 70 & 5 & $\cdots$ & 1 \\
	& Closed Dialogue & 120 & 61 & 11 & 231 & $\cdots$ & 0 \\
	& $\vdots$ & $\vdots$ & $\vdots$ & $\vdots$ & $\vdots$ & $\ddots$ & $\vdots$ \\
	& Not Interacting & 20 & 0 & 23 & 1 & $\cdots$ & 1647 \\
	\hline\hline
	\end{tabular}
\end{table*}
The diagonal of this matrix shows when the observations agree, with off-diagonals showing disagreement. Thus, to compute the number of agreements between two observations, one would simply compute the sum of the values along the diagonal. \hyperref[eq:agree]{Equation~\ref*{eq:agree}} shows this calculation for $n$ interaction types or an $n \times n$ matrix.

\begin{align}
	N_\mathrm{agreement} = \sum\limits_i^n a_{ii}
	\label{eq:agree}
\end{align}

The percent agreement would then be this value divided by the sum of all the values in the matrix. Cohen's kappa goes a step further to exclude from the agreement any paired observations that may occur due to chance. To determine the number of agreements that may occur due to chance, one would compute the product of total number of times each observer marked a given category, summing over all categories. Dividing this sum by the total of all values in the matrix yields the number of of chance agreements.

\begin{align}
	N_\mathrm{chance} = \frac{\sum\limits_i^n\left(\sum\limits_j^n a_{ij}\sum\limits_j^n a_{ji}\right)}{\sum\limits_i^n\sum\limits_j^n a_{ij}}
	\label{eq:chance}
\end{align}

Cohen's kappa is then determined by the difference of the numbers of agreement and chance divided by the difference of the total number of interactions and the number of chance agreements.

\begin{align}
	\kappa = \frac{N_\mathrm{agreement} - N_\mathrm{chance}}{\sum\limits_i^n\sum\limits_j^n a_{ij} - N_\mathrm{chance}}
	\label{eq:kappa}
\end{align}

It is important to remember that measurements of human interactions is inherently ``noisier'' than standard measurements in a physics laboratory environment. Values for Cohen's kappa range from $-1$ to $1$, with 1 indicating perfect agreement and a value of 0 indicating chance agreement. Arbitrary guidelines have been suggested to categorize kappa values from 0.61 to 0.80 as substantial and 0.81 to 1 as almost perfect agreement \cite{landisandkoch}. Another equally arbitrary guideline characterizes kappas over 0.75 as excellent and 0.40 to 0.75 as fair to good \cite{fleiss}. By either of these standards, my value of 0.825 is more than acceptable.

\chapter{A breakdown of the interactions used overall}
	\label{sec:overall_results}

\begin{figure*}[bt]
	\centering
	\includegraphics[width=\linewidth]{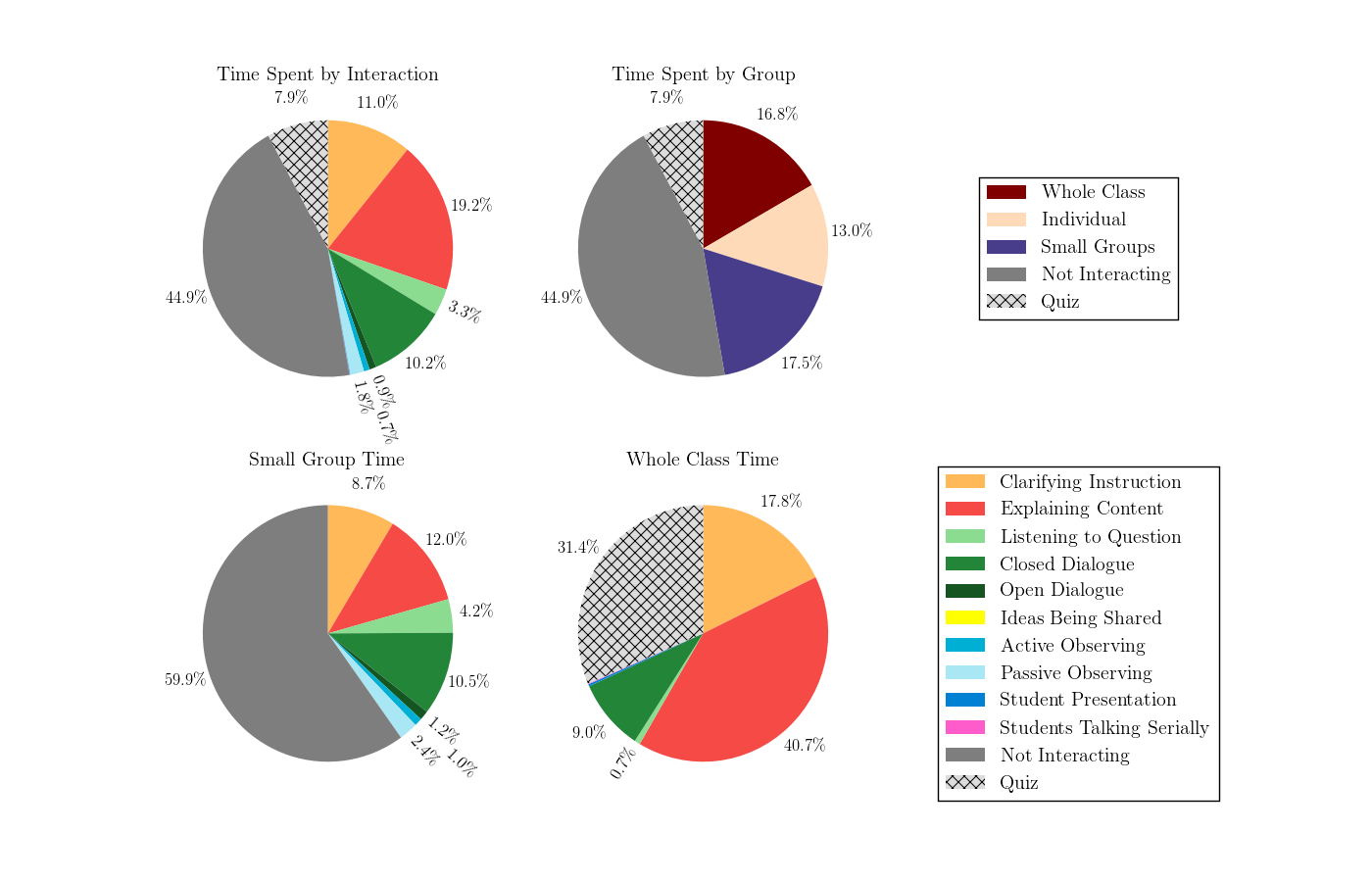}
	\caption[TA-student interactions in four different courses]{A group of pie charts showing the net results of over twenty hours of TA-student interactions throughout four different courses: Physics~2A Lab, Physics~50 Lab, Physics~50 Workshop, and Physics~51 Lab.}
	\label{fig:OverallPies}
\end{figure*}

Ultimately, I made observations in four different courses: Physics~50 and 50W (described in \hyperref[sec:sample]{Section~\ref*{sec:sample}}) as well as the algebra-based freshman mechanics lab (Physics~2A) and the calculus-based freshmen electricity and magnetism lab (Physics~51). \hyperref[fig:OverallPies]{Figure~\ref*{fig:OverallPies}} shows the net result of these observations. The four pie charts shows the sum of the interactions for ``Small Group'' time, ``Whole Class'' time, ``Small Group and Whole Class'' time, and a breakdown of ``Time Spent by Group.''

Looking at the sum total of my observations, the TAs spent just under a third of the time talking at their students: either {\em Clarifying Instruction} or {\em Explaining Content}. Less than a fifth of the total time was spent talking with or observing their students. TAs spent nearly 45\% of the total time {\em Not Interacting} with their students.

For my observations, I found it useful to note when the TAs administered a formal quiz. I used the comment section of the RIOT to note when the quiz began and ended. Later, I manually went through the csv file for each observation and manually changed the codes between the times noted by my comments to a new {\em Quiz} code. TAs and students had few interactions during this time, with individuals occasionally asking questions. While {\em Not Interacting} may have been an accurate description of what was happening during this time, it seemed unfair to weight this time as {\em Not Interacting} when {\em Quiz} was more fitting.

TAs tended to use this time to grade, prepare notes on the whiteboard for a lecture on the coming lab, and a negligible amount of passive observation. Using the original codes instead of my own {\em Quiz} code on the ``SG \& WC Time'' chart, the {\em Not interacting} time would increase by 7.6\%, with the remaining 0.3\% divided between {\em Clarifying instruction}, {\em Explaining Content}, and {\em Listening to Question}.

\chapter{Comparison of Physics~50 and Physics~50W}
	\section{Physics~50}
\begin{figure*}[tbh]
	\centering
	\includegraphics[width=\linewidth]{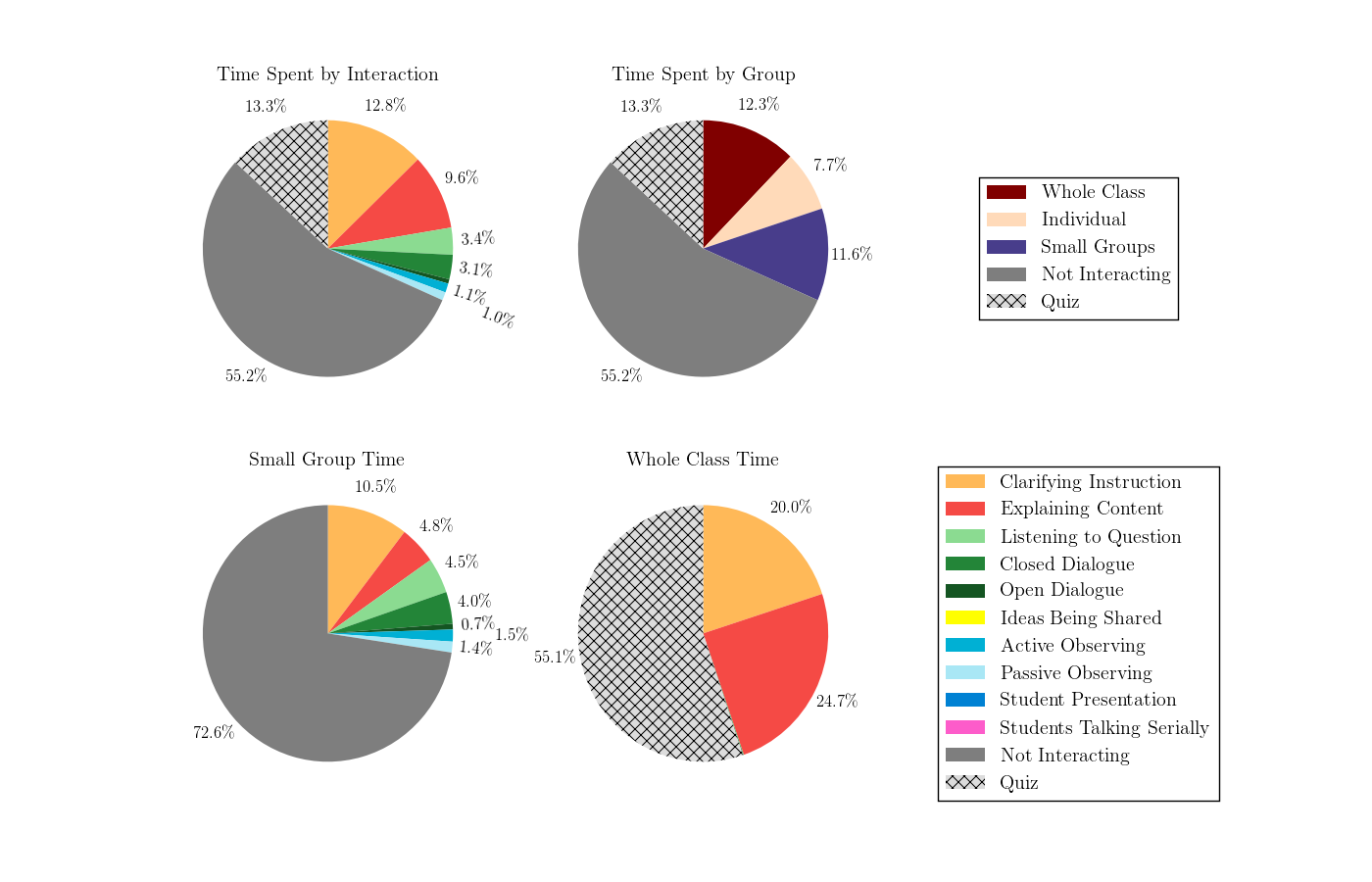}
	\caption[TA-student interactions in the Physics~50 lab classroom]{The net result of over nine hours of observations in the Physics~50 lab classroom. The percentages listed are the totals of all observations rather than averages.}
	\label{fig:50Pies}
\end{figure*}

		I made a total of nine observations in Physics~50 lab classes, visiting three different TAs as shown in \hyperref[tab:obs_counts]{Table~\ref*{tab:obs_counts}}. Of these nine, four began the meeting with a ``Proficiency Quiz.'' The quizzes lasted $20\pm2$~minutes and were followed by a brief lecture about the day's lab experiment or problem solving session. The lectures ranged from roughly 4 to 13~minutes, averaging $7\pm3$~minutes. As shown in the ``Whole Class Time'' pie chart in \hyperref[fig:50Pies]{Figure~\ref*{fig:50Pies}}, these lectures consisted almost entirely of {\em Explaining Content} and {\em Clarifying Instructions}. There exists a very small slice of green on that chart allowing for only a second or two of questions or dialogue with the students in average class.

How TAs allocated their lab time is broken down by group in the bottom right pie chart of \hyperref[fig:50Pies]{Figure~\ref*{fig:50Pies}}. We find 12.3\% of the total time observing was spent with the lecturing at the whole class. 7.7\% of the time was spent interacting with individual students and 11.6\% was spent interacting with students in small groups. Over half my total observation time in Physics~50 was spent with TAs {\em Not Interacting} with their students. I should note here that the RIOT considers {\em Observing} an interaction. This means 55.2\% of the time I spent observing the Physics~50 labs, the attention of the TAs was somewhere other than their students. Further, since I only observed the first hour of lab time, it is unlikely that students were working on completing their lab write ups, but rather actively collecting and analyzing data or working on Physics problems. I operated on the assumption that the TAs would be the most interactive during their first hour of teaching and that students would be finishing up their lab reports during the latter portion of lab time. Because of this, I expect that my results may show the TAs as more interactive than they actually are, if one were to observe them for the entirety of the class time.

\begin{figure*}[bt]
	\centering
	\includegraphics[width=0.32\linewidth]{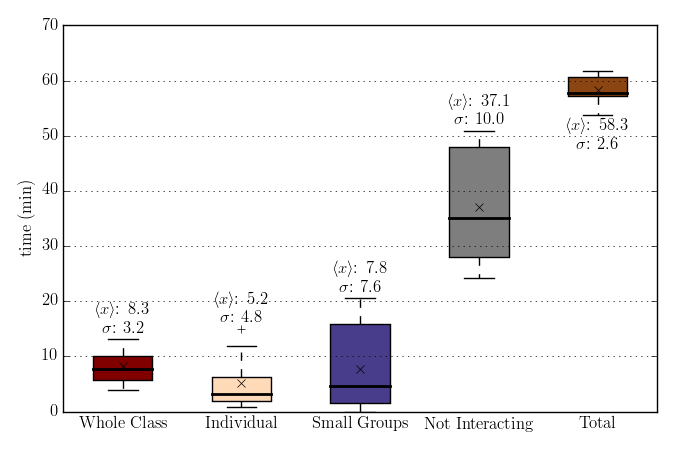}
	\includegraphics[width=0.32\linewidth]{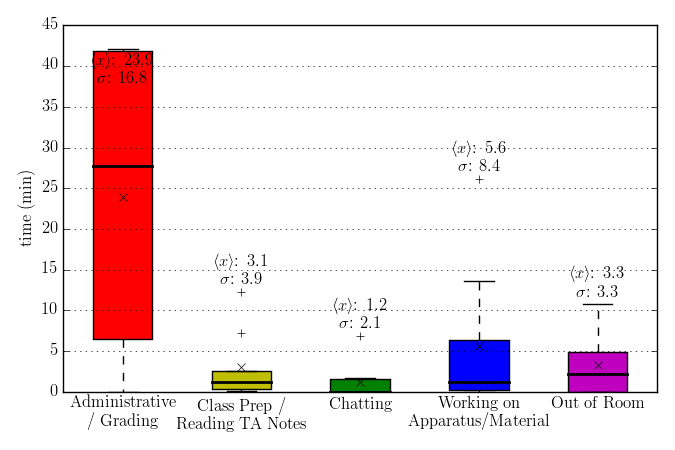}
	\includegraphics[width=0.32\linewidth]{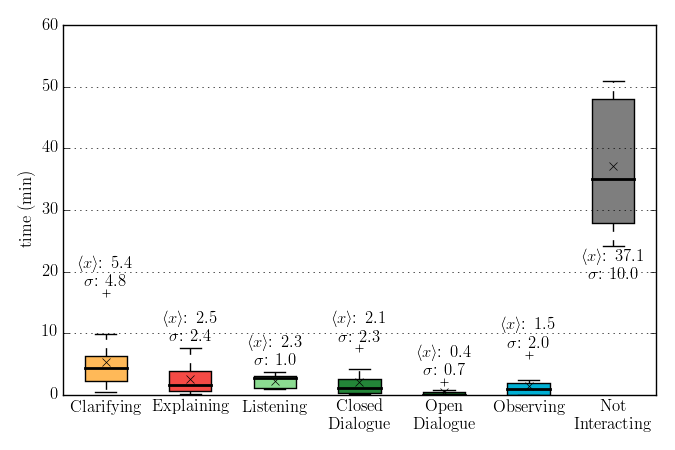}
	(a) Time spent by Group Type \hspace{0.05in}
	(b) {\em Not Interacting} \hspace{0.25in}
	(c) ``Small Group Time''
	\caption[Box-and-whisker plots of Physics~50 interactions]{The breakdowns of time spent between ``Whole Class Discussion'' and ``Small Group Time'' (left), {\em Not Interacting} (center), and ``Small Group Time'' (right) in the nine observations of Physics~50 labs. 
		The heavy bar across the middle of each box shows the median value for that category while the $\times$~marks the mean value and outliers are marked with $+$~signs. Mean values and standard deviations for each category are also printed. All numeric values are printed in minutes.
		}
	\label{fig:50boxplot}
\end{figure*}

It is perhaps more useful to examine how TAs allocated their time in terms of actual lab time. 
	\hyperref[fig:50boxplot]{Figure~\ref*{fig:50boxplot}}
	 shows the breakdown of the nine observations of Physics~50 labs, allowing for variation between observations.  Here we see that after the quiz and introductory lecture, TAs spent \unit[14]{minutes} of the \unit[51]{minutes} remaining in my observation interacting with their students.

With over 70\% of ``Small Group'' Time spent {\em Not Interacting}, it is important to breakdown this category into its constituent parts, as shown in 
	\hyperref[fig:50boxplot]{Figure~\ref*{fig:50boxplot}b}. In this plot, we see that much of the time is spent in the {\em Administrative/Grading} category. A portion of this time can be attributed to TAs grading Proficiency Quizzes and entering grades into books or computers. Grading was shortly followed by TAs {\em Out of Room} to scan the graded quizzes on the computer in the adjoining prep room. 
The total amount of time spent {\em Out of Room} is as expected and unremarkable. 

{\em Class Prep} time is reasonably low. This too can be partly attributed to the Proficiency Quizzes. TAs tended to use {\em Quiz} time to prepare their lecture notes on the whiteboard. The {\em Working on Apparatus/Material} category appears to be reasonable at first glance. It should be pointed out that two or three of the nine observations took place during ``Problem Solving Sessions.'' Since TAs are provided complete solutions to the problems and there is no equipment to malfunction during these meetings, it is possible that these sessions weigh the category lower than it would be for a pure laboratory observations.

It should be noted that the Proficiency Quizzes are designed to be quick and easy to grade, consisting of a single problem with upwards of four parts. TAs are provided with complete solutions. Judging by my observation notes, quiz grading can account for roughly five minutes the {\em Administrative/Grading} category. 

After dissecting {\em Not Interacting} time, it is also instructive to examine the breakdown of ``Small Group Time'' in terms of actual lab time, as shown in 
	\hyperref[fig:50boxplot]{Figure~\ref*{fig:50boxplot}c}. Here we see that, on average, TAs spent 7.9~minutes talking at students ({\em Clarifying Instructions} and {\em Explaining Content}). On the other hand, they spent only 4.8~minutes talking with students ({\em Listening to Questions} and {\em Closed} or {\em Open Dialogue}) and only 1.5~minutes {\em Observing} students working. 
The standard deviations in all the times spent interacting with students are comparable to the mean values.
	
	\section{Physics~50W}
		\begin{figure*}[ht]
	\centering
	\includegraphics[width=\linewidth]{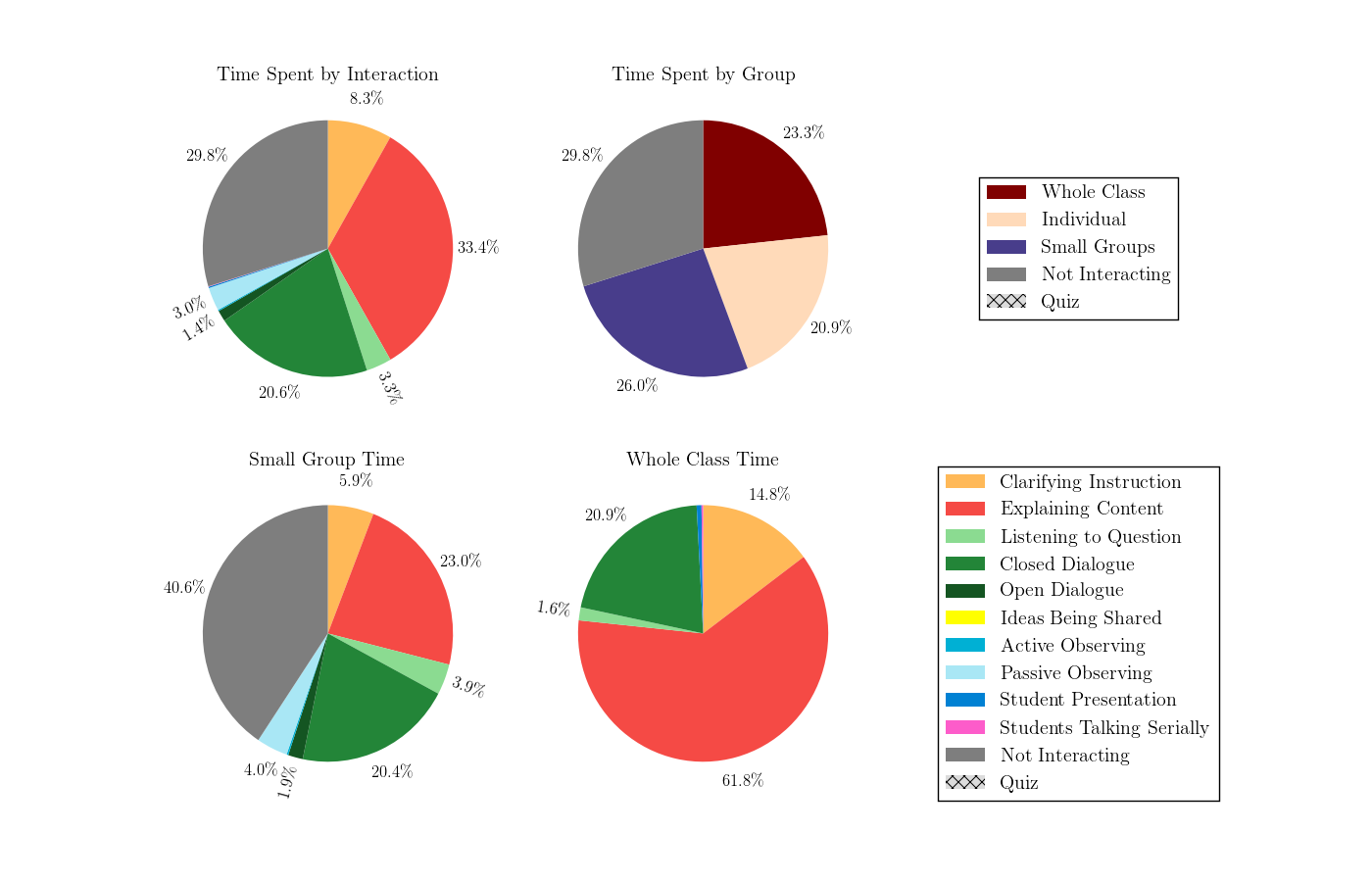}
	\caption[TA-student interactions in the Physics~50W classroom]{The net result of roughly seven hours of observations in the Physics~50 Workshops. The percentages listed are the totals of all observations rather than averages.}
	\label{fig:50WPies}
\end{figure*}

I made a total of seven observations in Physics~50 Workshop classes, visiting three different TAs as shown in \hyperref[tab:obs_counts]{Table~\ref*{tab:obs_counts}}. Students in these classes saw a wider range of interactions than those in the lab class. In fact, the pie charts in \hyperref[fig:50WPies]{Figure~\ref*{fig:50WPies}} contain all of the RIOT interactions to some extent except {\em Ideas Being Shared}. Starting with the ``Time Spent by Group'' pie chart, it seems that when viewed as a whole TAs split their time nearly evenly between working with the whole class, individual students, small groups, and not interacting. The {\em Not Interacting} portion is 29.8\% of the total time, followed by {\em Small Groups}, {\em Whole Class}, and {\em Individual} at 26.0\%, 23.3\%, and 20.9\%, respectively.

\begin{figure*}[tb]
	\centering
	\includegraphics[width=0.32\linewidth]{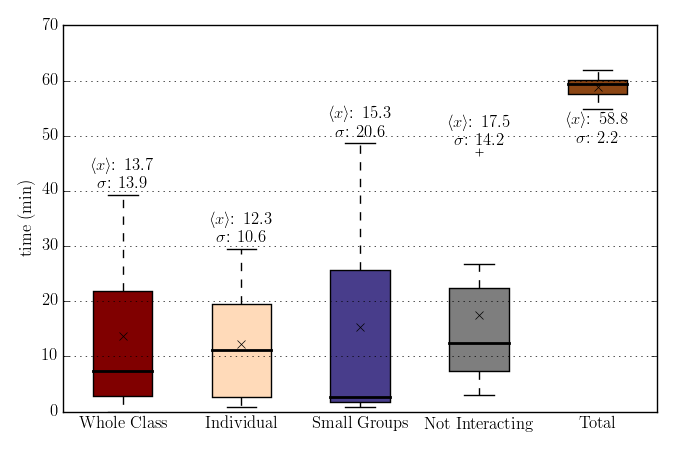}
	\includegraphics[width=0.32\linewidth]{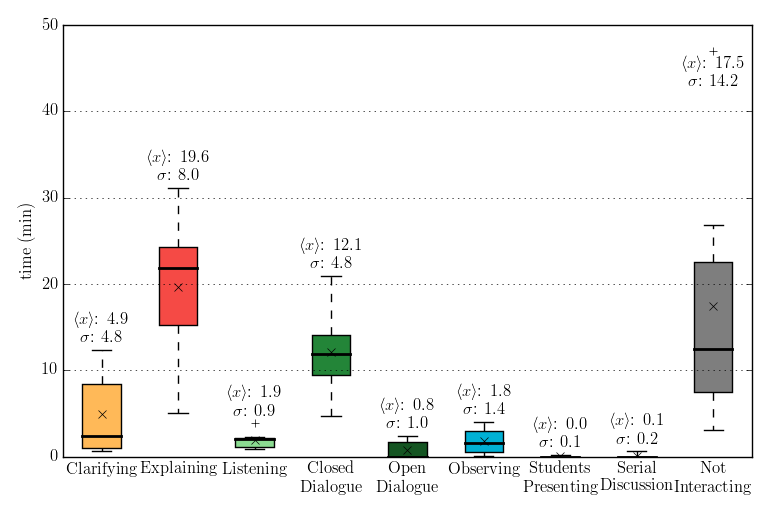}
	\includegraphics[width=0.32\linewidth]{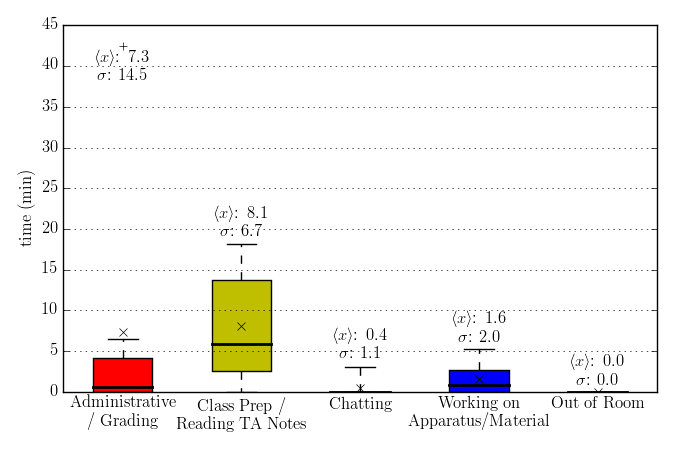}\\
	\hspace{-0.2in}
	(a) Time spent by Group Type \hspace{0.05in}
	(b) Time spent by Interaction \hspace{0.1in}
	(c) {\em Not Interacting}
	\caption[Box-and-whisker plots of Physics~50W interactions]{The breakdowns of time spent between ``Whole Class Discussion'' and ``Small Group Time'' (left), by interaction time in both ``Whole Class'' and ``Small Group'' (center), and {\em Not Interacting} (right) in the seven observations of Physics~50 Workshops. 
		The heavy bar across the middle of each box shows the median value for that category while the $\times$~marks the mean value and outliers are marked with $+$~signs. Mean values and standard deviations for each category are also printed. All numeric values are printed in minutes.
		}
	\label{fig:50Wbreakdown}
\end{figure*}

Not only was there a wider range of interactions for the workshops, but how TAs divided up their time in class was also more widely varied, as shown by the larger spreads in \hyperref[fig:50Wbreakdown]{Figure~\ref*{fig:50Wbreakdown}a}. Where one TA spent the majority of his class time interacting with the ``Whole Class,'' another TA spent none, going so far as not even greeting the class at the beginning of the session. Of the two TAs who did interact with the ``Whole Class,'' they both started each day with a short talk discussing the topic of the day. The format of the talk varied by day and by TA ({\em Clarify}, {\em Explain}, and {\em Dialogue} were all used) and ranged from two to eight minutes.

\hyperref[fig:50Wbreakdown]{Figure~\ref{fig:50Wbreakdown}b} shows the spreads of each interaction in both ``Whole Class'' and ``Small Group'' time. Since there is such a large variation in how TAs split their time between ``Whole Class'' and ``Small Group,'' it makes more sense in this case to ignore this distinction and look at how they interacted with their students regardless of setting. 

At first glance, there appeared to be far more {\em Clarifying Instruction} than I was expecting. Reading through the time series graphs and my notes for each observation, it became evident that the majority of this time was due to a single observation early in the semester. During that time, the TA was discussing the goals of the course and how it would be run, as per the syllabus. The rest came as incidentals, but more so from one TA than the other two.

The timing of the first ``Small Group'' interactions varied slightly by TA. One TA began interacting with her groups immediately following the session's introductory lecture, roughly ten to fifteen minutes in. She employed some {\em Active Observing} and roughly equal parts {\em Closed Dialogue} and {\em Explaining Content}. The other two TAs typically began interacting with their students twenty-five to thirty minutes into their sessions and even then generally only interacted with students individually. They employed short spots of {\em Observation} and tended {\em Explain Content} only slightly more often than {\em Dialoguing} with their students. A single TA spent much of each session {\em Explaining Content} to the ``Whole Class.'' He tended to spend more of his time discussing how to solve the problems on the worksheet than he did allowing his students time to work on them.

\begin{figure}[htb]
	\centering
	\includegraphics[width=3.5in]{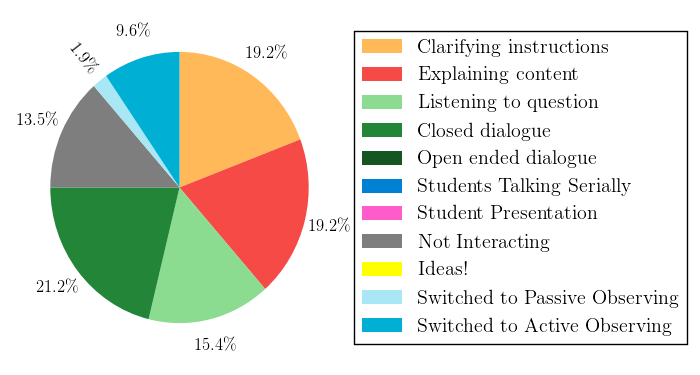}
	\caption[Interactions that directly followed {\em Observations} in Physics~50W]{Interactions that directly followed both {\em Passive} and {\em Active Observations} in Physics~50W.}
	\label{fig:50WfollowingObs}
\end{figure}

Slightly more {\em Observing} took place in the workshop setting than in the traditional lab. I wrote a quick python script to look through the time series data to see what interactions tended to follow these observations, the results of which are shown in  \hyperref[fig:50WfollowingObs]{Figure~\ref*{fig:50WfollowingObs}}. {\em Observing} led to {\em Closed Dialogue}, {\em Clarifying Instruction}, {\em Explaining Content}, and {\em Listening to Questions} with nearly the same frequency. This means that when the TAs saw an error or student difficulty, they were just as likely to attempt to correct it directly as they were to include the students in the reasoning process.

Even in the workshop setting, TAs tended to spent a lot of time {\em Not Interacting} with their students: averaging between a quarter and a third of the session. \hyperref[fig:50Wbreakdown]{Figure~\ref{fig:50Wbreakdown}c} shows the breakdown of this time. With the exception of an outlier in {\em Administrative/Grading} that likely would have been better categorized as ``off task,'' TAs spent the majority of their time doing {\em Class Prep/Reading TA Notes}. Reading through my observation notes, this appears to be time the TA spent reading through the provided solution set. The remaining {\em Working on Apparatus/Material} is when the TA was likely working out steps in the solution set for themselves.

	\section{A look at ``problem solving'' days}
\begin{figure*}[htb]
	\centering
	\includegraphics[width=\linewidth]{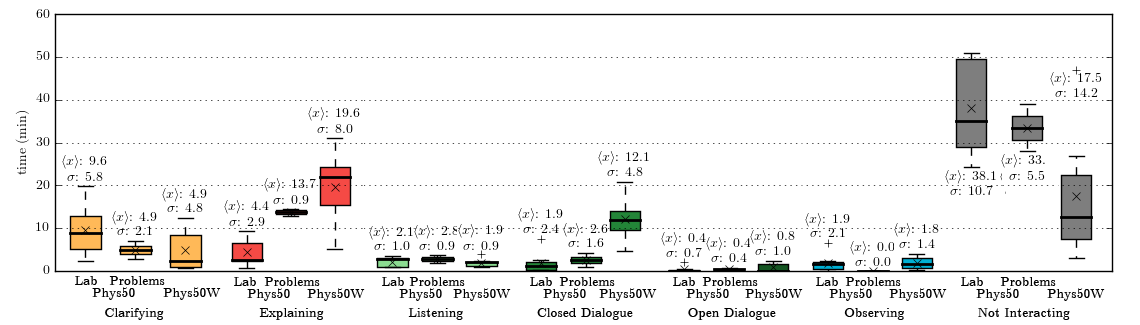}
	\caption[Box-and-whisker plot comparing Physics~50 lab days, Physics~50 problem solving days, and Physics~50W]{Interactions on regular lab days and problem solving days in Physics~50 compared with Physics~50W}
	\label{fig:problemsolvingdays}
\end{figure*}

		Only two of my Physics~50 observations were of ``problem solving sessions'' as opposed to the standard lab sessions. Consequently, I cannot make any generalizations comparing these sessions to the regular sessions or even to a typical Physics~50W session -- especially considering that both of these observations were of the same TA. Even without generalizability, the comparison is an interesting one to look at. \hyperref[fig:problemsolvingdays]{Figure~\ref{fig:problemsolvingdays}} shows the majority of the interactions logged by the RIOT divided into three groups: Standard Physics~50 lab sessions, Physics~50 problem solving sessions, and Physics~50W sessions. The interactions {\em Students Presenting} and {\em Students Talking Serially} were excluded because the values for all three session types were essentially zero.

Over nearly all interactions, it would appear that the problem solving sessions are being run similar to lab sessions. With the exception of {\em Explaining Content}, all the problem solving session interactions fall within or very near the middle 50\% of the spreads for the lab sessions. {\em Explaining Content} on the problem solving sessions is above the upper limit of the regular lab sessions and in the lower portion of the Physics~50W sessions.

Interestingly, there was no {\em Observing} taking place during the problem solving sessions. Looking over the time series data for these two observations, it appears the TA gave a short lecture on the basic physics involved and a few problem solving strategies lasting seven to fourteen minutes and then proceeded to only interact with individual students when they approached with questions.

Having only observed one TA for two problem solving sessions, I seriously doubt these observations would hold for the group of TAs as a whole. It follows then that more data needs to be collected here before we can attempt any generalization.

\chapter{Comparison to UC~Davis}
	The body of research utilizing the RIOT is still in its infancy. As of yet, there exist few holistic studies examining graduate student TAs interacting with their students \cite{West2013Variation,
			wilcox}. Consequently, we have very few opportunities to compare data at our own institution to similar data at similar institutions.
			
A study at UC~Davis examined their model-based CLASP course \cite{West2013Variation,
			Potter:Sixteen
			}. The course is their  ``nominally calculus-based'' introductory physics course. The interactive portion of the course, called ``Discussion/Lab'' (DL), has students work together in small groups to answer a series of activity prompts followed by a whole class discussion of the major points of the activity and explore general implications of the phenomena. Instructors, largely graduate students and a few faculty, are coached to guide students toward correct usage of the models rather than provide immediate answers.  Effectively, the CLASP course is designed to be a more interactive hybrid of both the Physics~50 lab and workshop courses at SJSU. As such, we might expect some similarities in the spreads of interaction types.

\begin{figure*}[htb]
	\centering
	\includegraphics[width=0.32\linewidth]{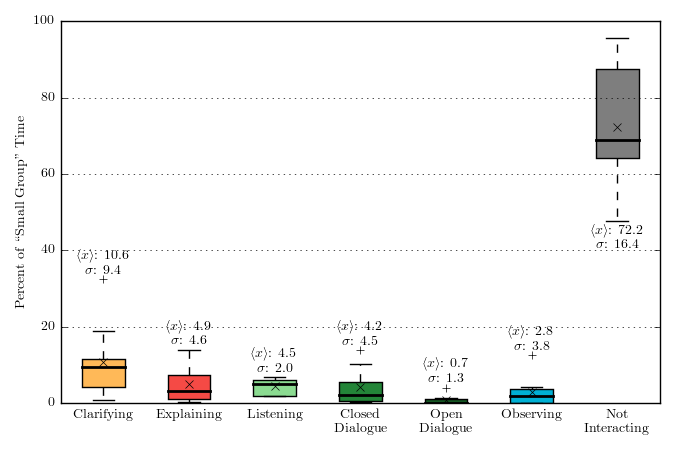}
	\includegraphics[width=0.32\linewidth]{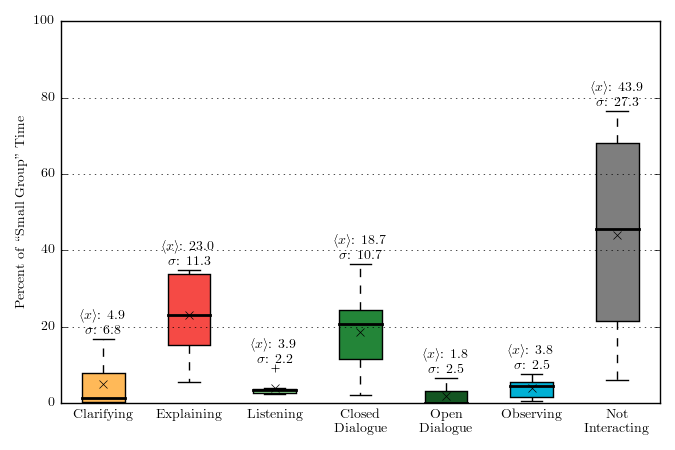}
	\includegraphics[width=0.32\linewidth]{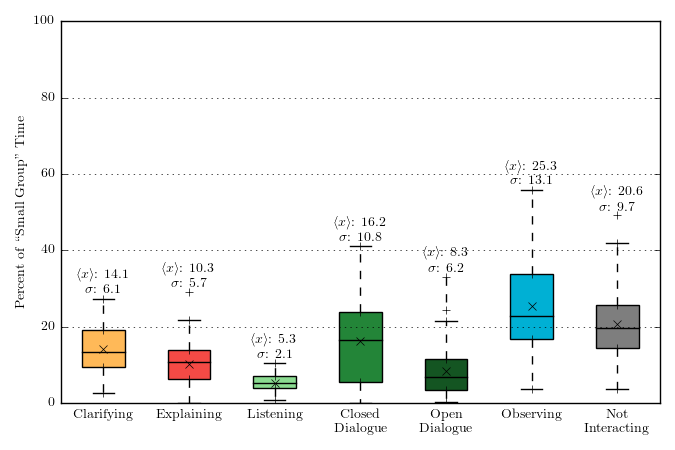}
	(a) SJSU Physics 50 \hspace{0.3in}
	(b) SJSU Physics 50W \hspace{0.3in}
	(c) UC Davis CLASP
	\caption[Box-and-whisker plot comparing ``Small Group'' time in SJSU and UC~Davis courses]{Breakdown of small group time by fraction of total small group time}
	\label{fig:CPSG}
\end{figure*}

\begin{figure*}[htb]
	\centering
	\includegraphics[width=0.32\linewidth]{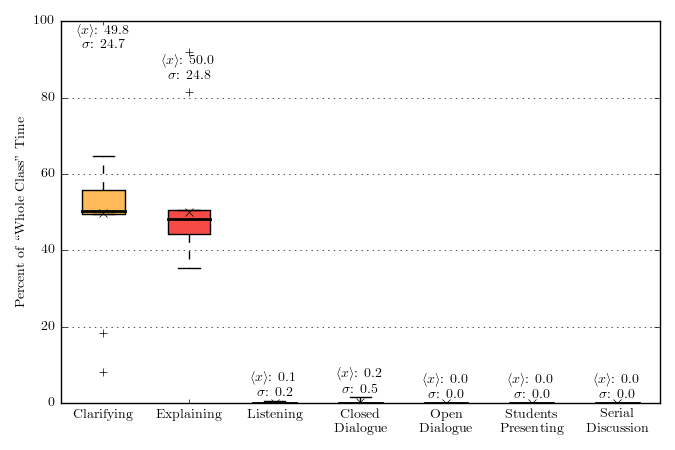}
	\includegraphics[width=0.32\linewidth]{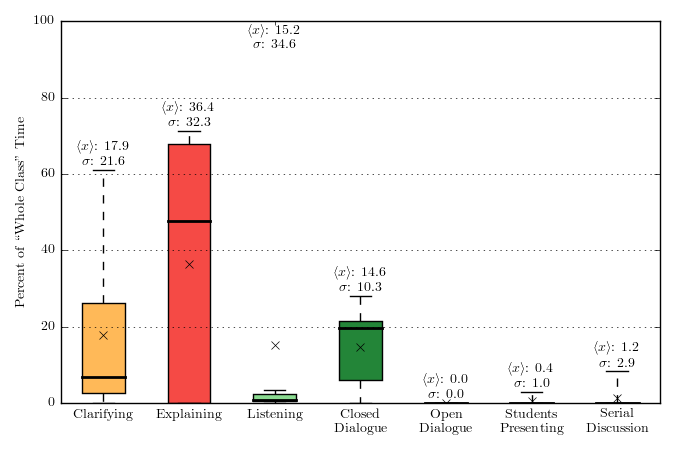}
	\includegraphics[width=0.32\linewidth]{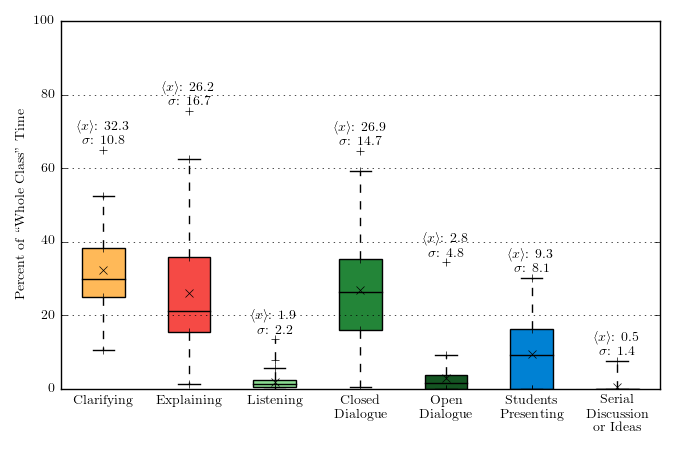}
	(a) SJSU Physics 50 \hspace{0.3in}
	(b) SJSU Physics 50W \hspace{0.3in}
	(c) UC Davis CLASP
	\caption[Box-and-whisker plot comparing ``Whole Class'' time in SJSU and UC~Davis courses]{Breakdown of whole class time by fraction of total whole class time}
	\label{fig:CPWC}
\end{figure*}

The observations from West, {\it et al.} lasted around 140~minutes whereas my own were less than half of that. The following comparisons are normalized by total time in a specific mode. For example, \hyperref[fig:CPSG]{Figure~\ref*{fig:CPSG}} shows the range of ``Small Group'' interactions normalized by the total amount of time spent in ``Small Group'' mode for each individual observation. \hyperref[fig:CPWC]{Figure~\ref*{fig:CPWC}} shows the range of ``Whole Class'' interactions normalized by the total amount of time each observation spent in ``Whole Class'' mode.

Looking at ``Small Group'' time, it is clear that for both of the SJSU courses, the percentage of time spent {\em Not Interacting} is much higher than the Davis course. The lower end of the workshop's {\em Not Interacting} is comparable to Davis's CLASP. It is reasonable then to conclude that with proper training and departmental support, TA's in Physics~50W could, at the very least, be as interactive as those in CLASP.

When the TAs at SJSU are interacting with their students, they tend to spend more time ``talking at'' their students than ``talking with'' them. While Physics~50W has a slightly higher median {\em Closed Dialogue} percentage, {\em Listening to Questions} is comparable across all courses. In both of the SJSU courses, the percentage of time spent {\em Observing} is far less than in the Davis course. Interestingly, Physics~50W TAs tended to spend a higher percentage of their ``Small Group'' time {\em Explaining Content} while Physics~50 lab TAs spent a lower percentage of their ``Small Group'' time {\em Explaining}. CLASP TAs fall neatly in between these two groups. They did tend to spend a larger fraction of this time {\em Clarifying Instruction} than TAs in both SJSU courses. This is likely due to the fact that the materials provided to SJSU students are far more explicit in their instructions for students compared to the DL materials provided to the CLASP students.

The use of ``Whole Class'' time varied significantly by course. With the exception of a small number of my Physics~50W observations, all courses began with a short talk from the TA to the ``Whole Class.'' For the Physics~50 labs, the interaction mode ended there as students broke up into their lab groups to conduct the experiment of the day. The same thing occurred for some of the TAs in the Physics 50W course. In the CLASP course, students frequently return to ``Whole Class'' mode to present and discuss worked problems and lab results. Looking at \hyperref[fig:CPWC]{Figure~\ref*{fig:CPWC}}, we again see only slight similarities between Physics~50W and CLASP. {\em Clarifying Instruction} and {\em Explaining Content} show similar ranges in both courses, likely due to variation in teaching style by TA. {\em Closed Dialogue} in Physics~50W is, at best, half that of CLASP while {\em Listening to Questions} is comparable for both courses.

The similarities end there. Students in both SJSU courses were never provided the opportunity to {\em Share their Ideas}, leading the conversation to collectively challenge their own ideas regarding physics concepts. They also never {\em Presented} their lab findings and rarely {\em Presented} to each other or {Talked Serially} to build each others' ideas. These interactions all show up in the CLASP observations, as {\em Ideas} and {\em Present.} While {\em Sharing Ideas} was mostly left to outliers, the opportunity for {\em Student Presentations} plays a fairly important role in the CLASP material and student sense-making in general.

Based on my observations, the Physics~50W course more closely aligns with UC Davis's CLASP course than the Physics~50 lab. Even so, the alignment 
shows more differences than similarities. While students in both courses at SJSU are engaged in their prescribed activities, they are rarely encouraged communicate their ideas through discussion with their peers or instructors, as evidenced by the small amount of time spent in {\em Dialogue}, even smaller amount of time spent {\em Presenting}, and the entire lack of {\em Sharing Ideas}. It seems fair to conclude that neither of the SJSU courses as they currently stand could fairly be classified as interactive courses.

TAs will require significantly more training and departmental support before either course will become more interactive. The data presented here illuminate many areas for potential growth. Future interventions and subsequent observations could easily be designed to target any specific deficiency already discussed.

\chapter{Alignment with PER-based best practices}
	The refrain common to much of PER is that IE classes are ``better.'' Students tend to achieve significantly higher gains on common conceptual tests like the {\em Force Concept Inventory} (FCI) \cite{fci}, the {\em Force and Motion Conceptual Evaluation} (FMCE), and the {\em Mechanics Baseline Test} (MBT) \cite{mbt}. A paper by Hake is frequently cited to show this is \cite{hake}. Looking closely at the data he provides reveals that there is a larger variation in the normalized FCI gains achieved in IE classes, ranging as low as the gains achieved in unreformed courses to much higher. Hake hypothesized that these variations may be due to variations in effectiveness of pedagogy and/or implementation. Subsequent research has examined implementations of a specific curriculum and found a  similar range of gains \cite{Pollock:Sustaining}.

	\section{Lessons from Peer Instruction}
		In seeking to learn more about what variations in teaching practices exist and what effect they may have on student learning, a team at the University of Colorado at Boulder observed six of their faculty teaching six different introductory physics courses \cite{Turpen2010Construction} using {\em Peer Instruction} (PI) \cite{mazur:PI}. Through their observations, they identified 13 different dimensions of practice (DoP) and measured significant variation in faculty application of these DoPs.

The DoPs are divided into two main groups, one for how the professor defines an academic task and another for student-professor interactions during the task. They noted little variation in how the six professors applied the first set of DoPs, citing similar choices in clicker question types (mostly conceptual questions) and approach to having students answer the clicker questions (allowing some quiet time for students to reflect on the question before discussing with their peers). They did however find significant variation in how the professors interacted with their students during the administration of the clicker questions. Only half of the professors ``left the stage'' while students worked on and discussed the questions. There was a large amount of variation in the professors answering student-initiated questions and engaging in discussion with the students. There was also a large variation in the amount of time allotted for students to answer clicker questions and how the answers to the clicker questions were discussed. Some instructors frequently asked for student explanations while other did so rarely. Similarly, some instructors discussed wrong answers occasionally while others did so rarely.

In analyzing their data, they identified a variety of scientific practices that they valued (shown in \hyperref[tab:PI]{Table~\ref*{tab:PI}}), noting that a student's opportunity to gain experience in any of these practices varied significantly by instructor. They noted that all classrooms offered students the opportunity to try out and apply new physical concepts as well as discuss physics content with their peers. They identified ``large discrepancies in students' opportunities to engage in formulating and asking questions, evaluating the correctness and completeness of problem solutions, interacting with physicists, identifying themselves as sources of solutions, explanations, or answers, and communicating scientific ideas in a public arena.''

\begin{table*}[hb]
	\centering
	\caption[Scientific practices emphasized through the use of Peer Instruction]{Scientific practices that students gain experience with through the use of Peer Instruction}
	\label{tab:PI}
	\begin{tabular}{rl}
	\hline\hline
		i.	&	Trying out and applying new physical concepts	\\
		ii.	&	Discussing physics content with their peers	\\
		iii.	&	Justifying their reasoning to their peers	\\
		iv.	&	Debating physical reasoning with their peers	\\
		v.	&	Formulating and asking questions	\\
		vi.	&	Evaluating the correctness and completeness of problem solutions	\\
		vii.	&	Interacting with physicists	\\
		viii.	&	Beginning to identify themselves as sources of solutions, explanations, or answers	\\
		ix.	&	Communicating in a public arena	\\
	\hline\hline
	\end{tabular}
\end{table*}

	\section{Looking at our practices}
		While some of the scientific practices outlined in \hyperref[tab:PI]{Table~\ref*{tab:PI}} are not directly measurable by the RIOT, TA-student interactions can shed light on others. TAs are spending little time with student questions and discussions, opting to talk at their students more often than with them. This leads to few opportunities for students to ``formulate and ask questions.'' Even in Physics~50W, students rarely presented their solutions to each other, leading to few opportunities for ``evaluating the correctness and completeness of problem solutions,'' ``beginning to identify themselves as sources of solutions, explanations, or answers,'' and ``communicating in a public arena.'' Additionally, my observations have shown that students are not offered as much time to ``interact with a physicist'' as we might have assumed.

If nothing else, my observations may serve to inform our department and those at a similar stage of development. It is easy to assume that laboratory courses and workshops/tutorials are intrinsically interactive. Without proper training as well as departmental and curricular support, TAs run the risk of defaulting to ``teaching how they were taught.'' As my data show, their teaching may lack the degree of interactivity desired.
		
	\section{Lessons from Modeling Discourse Management}
		In seeking a more student-centered classroom discourse, a two-year college teacher and PhD candidate at ASU developed Modeling Discourse Management for his PhD dissertation \cite{desbien}. Much of his work is not directly applicable to our labs and workshops in their current state, but there is something to gain from examining the components of his discourse management style. \hyperref[tab:desbien]{Table~\ref*{tab:desbien}} shows the components of Modeling Discourse Management.

\begin{table*}[tb]
	\centering
	\caption{Components of Modeling Discourse Management}
	\label{tab:desbien}
	\begin{tabular}{rl}
	\hline\hline
		$\bullet$	&	Deliberate creation of a cooperative learning community	\\
		$\bullet$	&	Explicit need for the creation of models in science (epistemology)	\\
		$\bullet$	&	Creation of shared inter-individual meaning	\\
		$\bullet$	&	Seeding	\\
		$\bullet$	&	Intentional lack of closure	\\
		$\bullet$	&	Inter-student discussion	\\
		$\bullet$	&	Formative evaluation	\\
	\hline\hline
	\end{tabular}
\end{table*}

Extensive reorganization of the courses and TA training would be required to address many of the components and, even still, the majority of these components are not directly measurable by the RIOT. Instead, we focus on those aspects that could be readily adopted into our own practices: seeding and inter-student discussion.

In a modeling classroom, new ideas are ``seeded'' by the instructor during small group time, often in the form of a question or a hint. Early in small group time, the instructor looks and listens for key words or pictures from the groups. The instructor then asks a leading question, or offers a hint, ``so the group has time to work out details and gain ownership of the seeded idea.'' Seeding can be used effectively to stimulate broader participation in the whole class discussion. Quieter groups or students can be drawn into the discussion by seeding them with important ideas or questions to ask the group as a whole. Desbien goes on to say that one of his goals for his classroom is for inter-student discussion to be the dominant form of discussion, saying that ``real discussions and real cognitive dissonance occur more frequently when students do not feel the pressure of an authority figure questioning their ideas.''

\chapter{Conclusions}

	\section{A Summary of Results}
		What follows is a short summary of my findings, organized by the appropriate research question. References to the original figures are included as well as a brief discussion of their implications and important points for future TA professional development.

\subsection{What TA-student interactions do our TAs employ in their teaching assignments?}

Looking at \hyperref[fig:OverallPies]{Figure~\ref*{fig:OverallPies}}, we can see that the TAs spent just over 30\% of the time ``talking at'' their students. About 17\% of the total time was spent ``talking with'' or observing their students. TAs spent 45\% of the total time {\em Not Interacting} with their students. They spent comparable amounts of time interacting with the Whole Class as they did with Small Groups and only slightly less time interacting with students Individually.

\subsection{How do these interactions differ in Physics~50 and Physics~50W?}

\hyperref[fig:50boxplot]{Figure~\ref*{fig:50boxplot}} shows relatively small spreads in the interactions used by TAs teaching Physics~50. There is a larger amount of spread in how they spent {\em Small Group} time and {\em Not Interacting} time. This is more likely due to variation by TA than any other factor. On the whole, TAs tended to spend nearly the same amount of time in any given interaction type, possibly due to similar thoughts and beliefs about how a lab course should be run.

\hyperref[fig:50Wbreakdown]{Figure~\ref*{fig:50Wbreakdown}} shows much greater variation, not only in how TAs divide their time by group type, but also by interaction type. There appears to be less of a consensus about how to run a workshop course. This leads to a varying student experiences in Physics~50W that depends on which TA leads their workshop. In the interest of a homogenized workshop experience, more training and departmental support is needed. Here, TAs could become more interactive with useful, clear-cut examples and guidance from more-experienced instructors. Observations of others teaching the same course would be beneficial to their professional development, as would time to reflect on their own teaching practices.

\renewcommand\thesubsubsection{\thesubsection.\alph{subsubsection}}
\subsubsection{How do Physics~50 ``problem solving days'' compare to regular lab days and Physics~50W observations?}

The results for the problem solving days appear to more closely align with Physics~50W than they do with lab days in Physics~50. The data set consists of only two observations of a single TA unfortunately. More data is needed before we can make any real conclusions.

\subsection{How do our TAs' implementations of Physics~50 and Physics~50W compare with available RIOT data from courses at UC~Davis?}

As seen in \hyperref[fig:CPSG]{Figures~\ref*{fig:CPSG}} and \hyperref[fig:CPWC]{\ref*{fig:CPWC}}, TAs at SJSU have a stronger tendency to ``talk at'' their students than their UC~Davis counterparts. They also spend more time {\em Not Interacting}. Our TAs will require significantly more training and curricular/departmental support to become more interactive. We should present them with professional development opportunities that emphasize the need for both {\em Observation} and {\em Dialogue} between TAs and their students.

\subsection{How do our TAs' implementations of Physics~50 and Physics~50W compare with PER-based best practices?}

Part of Hake's definition of Interactive Engagement called for ``immediate feedback through discussion with \ldots  instructors'' \cite{hake}. Looking at the results of my observations, it appears that our TAs may not be providing this immediate feedback due to the little amount of time spent {\em Observing} and the large amount of time spent {\em Not Interacting}. Koenig, {\em et al.} showed that students learn better working in small groups followed by {\em Dialogue} with a TA \cite{koenig}. Again, in preparing professional development for our TAs, we would do well to focus on the need for both {\em Observation} and {\em Dialogue} between TAs and their students.

	\section{Future Work}
		I expected to see a fair amount of {\em Not Interacting} time, but I was still somewhat surprised at the amount I saw. To combat this, Dr.~Paul and I planned a three hour ``boot camp'' for all TAs planning to teach in the semester of Fall 2015. For activities, the group compared and contrasted science and science teaching practices, read and discussed a paper from PER literature \cite{koenig}, had a brief overview of some PER data (including some of the data from this study), and watched and discussed a pair of videos designed for physics pedagogy courses \cite{periscope}. While that amount of time is by no means exhaustive, we tried to continually return the discussion to the importance of asking students questions to find out what they know as a means of interaction.

My main goal of the training session was to decrease the amount of time TAs spent not interacting with their students while simultaneously increasing the amount of time they spent both observing and engaging in dialogue. A simple follow up study  conducting RIOT-based observations this and next semester could be done to see if this occurred.

It may also be beneficial to both TAs and the department to continue these observations. My biggest disappointment with this study is that I was never observed. I imagine the data from a pair or even a single observation would be invaluable to the professional development of both new and experienced teachers. Further, the observations can continue to be used to guide the development of future TA training and pedagogy courses. These courses will be the most effective when they specifically target the needs as seen in the data.

\appendix

\references{unsrt}{Bibliography}

\end{document}